\documentclass[lettersize,journal]{IEEEtran}
\usepackage{amsmath,amsfonts}
\usepackage{algorithm}
\usepackage{algorithmicx}
\usepackage{multirow}
\usepackage{booktabs}
\usepackage{hyperref} 
\usepackage{algpseudocode}
\usepackage{tabularx}
\usepackage{array}
\usepackage[caption=false,font=normalsize,labelfont=sf,textfont=sf]{subfig}
\usepackage{textcomp}
\usepackage{cite}
\usepackage{stfloats}
\usepackage{url}
\usepackage{verbatim}
\usepackage{graphicx}
\begin{document}
\title{BiT-MamSleep: Bidirectional Temporal Mamba for EEG Sleep Staging}

\author{
    Xinliang~Zhou\textsuperscript{*},~\IEEEmembership{}
        Yuzhe~Han\textsuperscript{*},~\IEEEmembership{} 
        Zhisheng Chen,~\IEEEmembership{}
        Chenyu~Liu,~\IEEEmembership{}
        Yi~Ding,~\IEEEmembership{}
        Ziyu~Jia,~\IEEEmembership{}
        and Yang~Liu~\IEEEmembership{}
        \thanks{{*} Xinliang Zhou and Yuzhe Han are with equal contributions.}
        \thanks{Xinliang Zhou, Chenyu Liu, Yi Ding and Yang Liu are with the College of Computing and Data Science, Nanyang Technological Univeristy, 50 Nanyang Avenue, 639798, Singapore (email: xinliang001@e.ntu.edu.sg, chenyu003@e.ntu.edu.sg, ding.yi@ntu.edu.sg and yangliu @ntu.edu.sg).}
        
        \thanks{Yuzhe Han is with the School of Electronic Engineering and Computer Science, Queen Mary University of London, Mile End Road, London E1 4NS, United Kingdom (email: jp2021212807@qmul.ac.uk).}
        \thanks{Zhisheng Chen is with the Beijing Key Laboratory of Mobile Computing and Pervasive Device, Institute of Computing Technology, Chinese Academy of Sciences, Beijing 100190, China (e-mail: zhishengchen@cumt.edu.cn).}
        \thanks{Ziyu Jia is with the Brainnetome Center, Institute of Automation, Chinese Academy of Sciences, Beijing 100190, China and also with the University of Chinese Academy of Sciences, Beijing 100190, China (email: jia.ziyu@outlook.com).}
        }
\markboth{IEEE INTERNET OF THINGS JOURNAL, VOL.}%
{Shell \MakeLowercase{\textit{et al.}}: A Sample Article Using IEEEtran.cls for IEEE Journals}



\maketitle

\begin{abstract}

In this paper, we address the challenges in automatic sleep stage classification, particularly the high computational cost, inadequate modeling of bidirectional temporal dependencies, and class imbalance issues faced by Transformer-based models. To address these limitations, we propose BiT-MamSleep, a novel architecture that integrates the Triple-Resolution Convolutional Neural Network (TRCNN) for efficient multi-scale feature extraction with the Bidirectional Mamba (BiMamba) mechanism, which models both short- and long-term temporal dependencies through bidirectional processing of electroencephalogram (EEG) data. Additionally, BiT-MamSleep incorporates an Adaptive Feature Recalibration (AFR) module and a temporal enhancement block to dynamically refine feature importance, optimizing classification accuracy without increasing computational complexity. To further improve robustness, we apply optimization techniques such as Focal Loss and SMOTE to mitigate class imbalance. Extensive experiments on four public datasets demonstrate that BiT-MamSleep significantly outperforms state-of-the-art methods, particularly in handling long EEG sequences and addressing class imbalance, leading to more accurate and scalable sleep stage classification.
\end{abstract}

\begin{IEEEkeywords}
Sleep Stage Classification, Electroencephalogram, Bidirectional Mamba, Class Imbalance, Temporal Dependencies, Computational Efficiency.
\end{IEEEkeywords}

\section{Introduction}
\IEEEPARstart{S}{leep} is a vital physiological process that plays a crucial role in maintaining health and well-being. Poor sleep is linked to various serious health conditions, including cardiovascular diseases, cognitive impairments, and mental health disorders \cite{ref1, eeg}. Understanding and classifying sleep stages is essential for diagnosing sleep disorders, which affect a significant portion of the population \cite{iot1, iot3}. Electroencephalogram (EEG) signals, which capture brain activity during sleep, serve as a critical tool for analyzing these stages, as they provide insight into the physiological and neurological patterns associated with different sleep phases. EEG-based sleep stage classification is thus crucial for enabling accurate diagnoses and guiding effective treatments.

Traditional sleep stage classification methods relied on expert manual annotation of EEG signals, which was time-consuming and labor-intensive. Early automated approaches such as support vector machines (SVMs) \cite{ref2,ref3} aimed to reduce human involvement, but their reliance on hand-crafted features limited their ability to generalize across different datasets and subjects. As deep learning advanced \cite{history}, convolutional neural networks (CNNs) and recurrent neural networks (RNNs) became prominent for sleep stage classification, enabling direct feature extraction from raw EEG signals . However, CNNs, with their fixed receptive fields, struggled to capture long-range temporal dependencies  \cite{ref8}, while RNNs, despite modeling temporal patterns, suffered from gradient issues when processing long EEG sequences \cite{ref15, CRbug}. These limitations underscored the need for models capable of efficiently capturing both short- and long-term dependencies in sleep data.

Transformers, with their self-attention mechanism, represent a major leap in modeling both short- and long-term dependencies without the vanishing gradient problem, making them highly effective for sequential tasks like sleep stage classification \cite{ref16, t2, ref17}. As a result, Transformer-based models have gained traction in the field of automatic sleep stage classification. However, these models face significant limitations, particularly when applied to the long, high-resolution EEG sequences typical in sleep studies.

The high computational cost of Transformers poses a significant obstacle when processing long sleep EEG sequences \cite{iot2}. Sleep EEG data, typically recorded at high sampling rates over extended durations, consist of a large number of time steps, resulting in highly complex datasets. Transformers, due to their quadratic increase in computational complexity as sequence length grows, struggle to efficiently process such data. This computational inefficiency poses a significant limitation, particularly in real-world sleep stage classification tasks where handling long EEG sequences with scalability is crucial. Therefore, it is crucial to develop models with reduced computational overhead and enhanced scalability for processing complex EEG data \cite{iot5, iot6, iot7, iot8}.

Capturing the complex bidirectional temporal dependencies in sleep stage transitions is another significant challenge for existing Transformer models. Sleep stage transitions involve subtle and gradual changes over time, requiring models to consider both past and future contexts. However, Transformers primarily operate in a unidirectional manner within fixed time windows, limiting their ability to model bidirectional dependencies. This limitation is particularly evident in long EEG sequences, where capturing the gradual transitions between stages like REM and NREM is critical. Due to their unidirectional nature, Transformers fail to fully capture these intricate transitions, which compromises their performance in accurately identifying sleep stage changes over extended periods. Therefore, models that can effectively capture bidirectional temporal dependencies are essential for handling long sleep EEG sequences.

Class imbalance further complicates sleep stage classification, highlighting another limitation of Transformer-based models \cite{iot9, iot10, iot11}. Sleep stages like REM occur far less frequently than others, such as NREM, leading to skewed predictions toward the majority classes. Transformers, lacking mechanisms to address this imbalance, tend to underperform when classifying minority stages. This imbalance negatively impacts overall classification accuracy, particularly in detecting underrepresented sleep stages. Effective handling of class imbalance is essential to improve the performance of models in sleep stage classification tasks, making it a key area for optimization in future approaches.

\begin{figure*}[!t]
\centering
\includegraphics[width=7in]{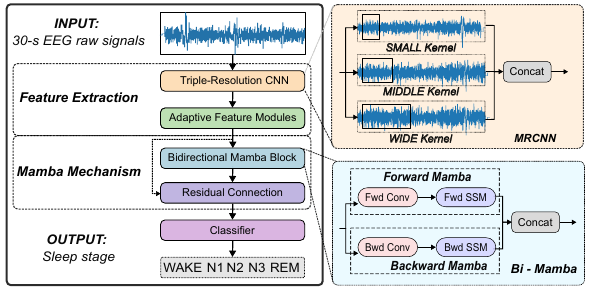}
\caption{An overview of the BiT-MamSleep architecture, which consists of two main functional components: the feature extraction module and the Mamba mechanism. The feature extraction module extracts relevant features from raw EEG signals. After mamba module which captures and focuses on the most salient information, the classification module, enhanced by custom imbalanced data handling techniques and dynamic learning rate adjustment strategies, outputs the predicted sleep stage.}
\label{fig_1}
\end{figure*}

To address these challenges, we integrate the feature extraction capabilities of the Triple-Resolution Convolutional Neural Network (TRCNN) with the Bidirectional State Space Model (BSSM). The TRCNN efficiently captures multi-scale frequency features essential for accurate sleep stage identification, while the Bidirectional Mamba (BiMamba) compensates for Transformer limitations by modeling both short- asnd long-term temporal dependencies through bidirectional EEG data processing. We further enhance model performance by introducing an Adaptive Feature Recalibration (AFR) module, which dynamically refine feature importance, optimizing classification accuracy without increasing computational complexity. Additionally, we propose a series of data optimization techniques to mitigate class imbalance and improve model stability during training, ensuring robust generalization across diverse sleep datasets.

Overall, the main contributions of our proposed model can be summarized as follows.
\begin{enumerate}
\item{We propose BiT-MamSleep, a novel architecture that integrates the TRCNN with the BiMamba mechanism. This design efficiently captures multi-scale spatial and temporal features from EEG data while maintaining linear computational complexity, allowing for scalable and efficient processing of long EEG sequences.}
\item{Bidirectional Temporal Modeling: The BiMamba mechanism utilizes BSSM which captures both past and future temporal dependencies, improving the accuracy of detecting transitions}
\item{Handling Data Imbalance: We address data imbalance and optimization challenges using Focal Loss and SMOTE, improving model generalization by focusing on difficult-to-classify samples and generating synthetic minority class data.}
\item{Extensive experiments on three public datasets show that our architecture outperforms state-of-the-art methods, especially in handling long sequences and imbalanced data.}
\end{enumerate}

Section II presents a comprehensive review of related work in the domain of sleep stage classification. Section III details the proposed BiT-MamSleep model, including the feature extraction module, the Mamba mechanism, and optimization techniques such as the class-aware loss function and learning rate adjustment strategies. Section IV introduces the datasets, evaluation metrics, and experimental setup, followed by a performance comparison with baseline models and an ablation study to analyze the model's components. Finally, Section V concludes the paper by summarizing key findings and proposing directions for future research.

\section{Related Work}

\subsection{Early approaches of automatic sleep stage classification}

Traditional sleep stage classification methods relied on expert manual annotation of EEG signals, which was time-consuming and labor-intensive. As sleep studies expanded, manual annotation became impractical, prompting the development of automated systems using machine learning techniques, such as SVMs and other models \cite{EDF_pre2}. While these methods reduced human involvement by automating feature extraction, they depended on hand-crafted features, which limited their generalizability across different datasets, subjects, and recording conditions. Variations in EEG recordings between individuals posed challenges to model robustness, and these models often struggled to maintain accuracy when confronted with new, unseen data. As more diverse EEG data became available, the limitations of feature engineering became evident, leading to the adoption of more flexible and powerful approaches such as deep learning.

\subsection{Deep Learning Models: CNNs and RNNs}

The introduction of deep learning marked a significant shift in sleep stage classification, particularly through the use of CNNs \cite{ref10,ref11}. CNNs demonstrated effectiveness in capturing local spatial features from raw EEG signals, eliminating the need for manual feature engineering. However, CNNs' fixed receptive fields limit their ability to model long-term dependencies, which are crucial for detecting transitions between sleep stages like REM and NREM. RNNs, designed to handle sequential data, offered an alternative for modeling temporal dependencies \cite{ref12,ref13,ref14}. Despite their advantages, RNNs faced the well-documented vanishing and exploding gradient problems when dealing with long EEG sequences. These issues hindered RNNs' effectiveness in accurately capturing long-term dependencies, revealing the need for a model that could efficiently handle both short- and long-term dependencies in sleep data.

\subsection{Transformer-Based Models}

Transformer models, with their self-attention mechanisms, offered a substantial improvement over CNNs and RNNs by effectively capturing both short- and long-term dependencies, without suffering from the vanishing gradient issues typical of RNNs \cite{need, layer}. This made them a popular choice for sequential tasks like sleep stage classification. However, Transformers face significant limitations when processing long EEG sequences, particularly due to their inability to model dependencies beyond a finite window. The quadratic scaling of computational costs with window length makes Transformers computationally inefficient and resource-intensive, especially when dealing with large, high-resolution EEG datasets. Additionally, their unidirectional processing of temporal dependencies constrains their ability to fully capture the bidirectional temporal relationships that are essential for accurately identifying sleep stage transitions over longer periods.

\subsection{State Space Models (SSMs) and the Mamba Mechanism}
State Space Models (SSMs) \cite{SSM1, SSM3, SSM2} provide an efficient alternative to Transformers by capturing complex temporal dependencies with linear computational complexity, making them suitable for processing long-duration EEG sequences. Mamba further optimizes SSMs through a hardware-aware design and dynamic time-step selection strategy, improving computational efficiency and ensuring the model focuses on the most relevant EEG segments \cite{mamba}. However, both SSMs and Mamba still struggle with capturing bidirectional dependencies, which are critical for understanding sleep stage transitions.

The BiMamba mechanism was specifically developed to overcome these limitations  \cite{visionmamba}. By processing EEG data bidirectionally, BiMamba simultaneously captures past and future information, allowing for a more accurate representation of sleep stage transitions, particularly in long sequences. This dual processing mechanism significantly improves the accuracy of sleep stage classification without increasing computational overhead, making it a superior solution for addressing the limitations of Transformers and traditional state space models. The integration of BiMamba into our proposed model allows it to effectively capture both short- and long-term dependencies, providing a scalable and efficient solution for sleep stage classification.

\section{Methodology}

The architecture of BiT-MamSleep consists of two main functional modules, as shown in Fig. 1. The first component is the feature extraction module, which is designed to capture both fine-grained and coarse-grained features from raw EEG signals across various temporal and frequency scales. The second component is the Mamba mechanism, which effectively models temporal dependencies by focusing on the most salient features through forward and backward sequence modeling. After the feature extraction and refinement, the model proceeds to predict sleep stages based on the processed features. Additionally, we have designed a series of Optimization Techniques and Data Handling methods to improve both the robustness and efficiency of the model. The following subsections provide a detailed explanation of each module.

\subsection{Feature Extraction}
The Feature Extraction Module is designed to effectively extract multi-scale and multi-frequency features from raw EEG signals, thereby enhancing the accuracy of sleep stage classification. Sleep stage classification relies on extracting frequency-specific features from EEG signals, as different sleep stages are characterized by distinct frequency bands \cite{ref14}. To achieve this goal, the Feature Extraction Module consists of two main components. First, the Multi-Resolution Convolutional Neural Network (MRCNN) is employed to capture rich temporal and frequency-domain features at various scales. Then, the AFR module dynamically adjusts and optimizes these features by learning the interdependencies between them. Below, we provide a detailed explanation of each of these components, as illustrated in Fig. 2.

\subsubsection{Triple-Resolution Convolutional Neural Network} 
To effectively extract features from raw EEG signals, we developed a TRCNN with three distinct branches of convolutional layers. Each branch employs a unique kernel size to capture specific temporal and frequency characteristics of the EEG signals. The use of three branches comprehensively addresses the non-stationary nature of EEG signals, where different sleep stages exhibit distinct frequency bands, such as delta waves in deep sleep (N3) and alpha waves in light sleep (N1). Compared to dual-branch designs, this multi-resolution approach enables our model to capture a broader spectrum of frequency-specific information, thereby enhancing the feature representation for sleep stage classification.


The three branches in the TRCNN architecture are configured with different convolutional kernel sizes to target distinct frequency ranges:
\begin{itemize}
\item \textbf{Small Kernel Branch}: Uses a kernel size of 50 to capture high-frequency details, critical for identifying rapidly changing EEG patterns associated with alpha and beta waves.
\item \textbf{Medium Kernel Branch}: Uses a kernel size of 100 to capture intermediate frequencies, balancing local and global temporal information. This scale is essential for features that are neither purely local nor extremely global.
\item \textbf{Large Kernel Branch}: Uses a kernel size of 400 to capture low-frequency information, such as delta waves dominant during deep sleep stages, reflecting long-range dependencies.
\end{itemize}

Each branch (CNN-unit) is a fully convolutional architecture inspired by prior work \cite{ref19}, which reduces the number of parameters, enhances the feature representation in convolutional layers, and improves model generalization. It involves convolutional layers, batch normalization (BN) layers \cite{ref20}, GELU activation functions, max pooling layers, and dropout, as illustrated in Fig. 2. Convolutional layers extract specific frequency features, BN normalizes feature maps to stabilize training, and GELU activation allows some negative inputs, making feature extraction more flexible. Max pooling downsamples the feature maps, and dropout prevents overfitting. Each branch also includes a repeated "convolution-BN-activation-pooling" block, repeated twice, to deeply extract features.

\begin{figure}[!t]
\centering
\includegraphics[width=3.5in]{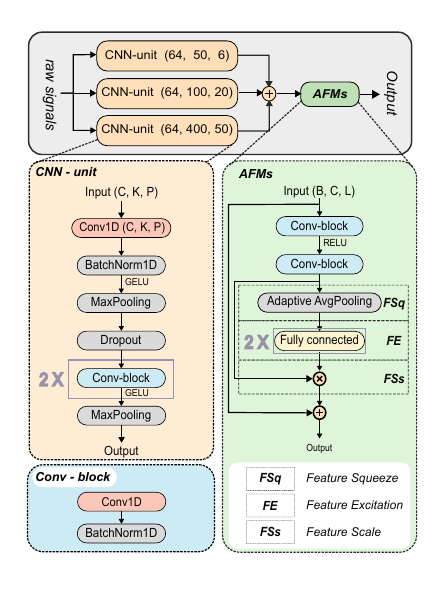}
\caption{The structure of Feature Extraction modules. Each CNN-unit is represented as (Out Channels, Kernel Size, Padding). The shaded areas represent layers that do not participate in parameter updates (e.g., Dropout layers). These layers do not adjust model weights during the training process.}
\label{fig_2}
\end{figure}
\subsubsection{Adaptive Feature Modules (AFMs)} 
To enhance feature extraction, we designed Adaptive Feature Modules, decomposing the original AFR into three submodules: Feature Squeeze (FSq), Feature Excitation (FE), and Feature Scale (FSs). This modular approach incrementally improves feature representation
\begin{itemize}
\item{Feature Squeeze (FSq): The main function of the Feature Squeeze module is to reduce feature redundancy and computational complexity while retaining key information. This is achieved by applying global average pooling to map multi-scale features into a more compact representation. Specifically, the global average pooling operation can be defined as follows:
\begin{equation}
z_c = \frac{1}{H \times W} \sum_{i=1}^H \sum_{j=1}^W x_{ijc}
\end{equation}

where $z_c$ is the output of the $c$-th channel, $H$ and $W$ are the height and width of the input feature map, and $x_{ijc}$ represents the feature value at position $(i, j)$ in the $c$-th channel. This pooling process helps streamline the input for the next module.This provides a compact yet informative representation to serve as input for subsequent modules. \cite{ref21}}
\item{FE: The Feature Excitation module utilizes a two-layer fully connected network, similar to the SE layer \cite{ref22}, to learn dependencies between different feature channels and adaptively adjust their importance to enhance the feature representation. The first layer reduces dimensionality using ReLU, and the second restores it using Sigmoid, allowing the model to focus on discriminative features, thereby improving classification accuracy.}
\item{FSs: The Feature Scale module combines the excited features from Feature Excitation to better capture inter-scale feature relationships. Through a channel-wise scaling operation, each channel of the input features is reweighted, emphasizing important features and suppressing irrelevant ones to enhance the final feature representation. A shortcut connection is added to retain original information and ensure stability \cite{ref23}}. 
\end{itemize}

These three modules work synergistically to make the feature extraction process more efficient and robust, significantly enhancing model performance, especially in handling high-dimensional and non-stationary EEG signals.While the AFR module enhances the local feature representation, capturing global temporal dependencies requires a more robust mechanism, which is where the Mamba Mechanism comes into play.

\begin{figure}[!t]
\centering
\includegraphics[width=3.5in]{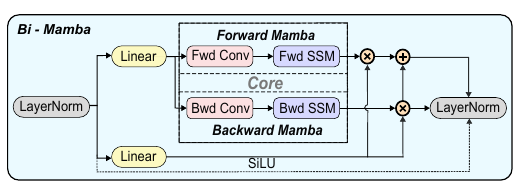}
\caption{The Mamba Mechanism distinguishes itself from conventional CNNs and RNNs by its ability to efficiently capture long-range temporal dependencies while minimizing computational overhead, a common limitation in traditional models.}
\label{fig_3}
\end{figure}

\subsection{Mamba Mechanism}
The Mamba Mechanism is designed to effectively capture temporal dependencies, thereby enhancing feature representation for improved sleep stage classification. In EEG signal processing, it is vital to capture both short-term and long-term temporal relationships, as sleep stages are defined not only by their distinct frequency components but also by their intricate temporal dependencies \cite{ref24}. To achieve this, the Mamba Mechanism incorporates the Bidirectional Mamba module, which models temporal dependencies in both forward and backward directions, resulting in a more comprehensive representation of EEG data while maintaining computational efficiency, as illustrated in Fig. 3.

\subsubsection{Bidirectional Mamba} 
The Bidirectional Mamba module is designed to effectively capture both short- and long-term temporal dependencies in EEG data, a critical task for accurate sleep stage classification. The module consists of two main stages: preprocessing via linear projection and layer normalization, followed by the core Bidirectional Mamba mechanism, as illustrated in Fig. 3.

The core of the Bidirectional Mamba involves a combination of convolutional layers and State Space Models (SSMs). For each direction—forward and backward—the convolution layer extracts local temporal features from the EEG sequence, while the SSM captures long-term temporal dependencies through discrete state updates. The state evolution is inspired by a continuous-time state space system described as:

\begin{equation} h'(t) = A h(t) + B x(t), \quad y(t) = C h(t)\end{equation}

Here, the input sequence $x(t) \in \mathbb{R}$ is processed through a hidden state $h(t) \in \mathbb{R}^N$, which evolves based on a state transition matrix $A \in \mathbb{R}^{N \times N}$ and an input matrix $B \in \mathbb{R}^{N \times 1}$. The output is generated through matrix $C \in \mathbb{R}^{1 \times N}$, providing a mapping from the hidden state.

To make this system applicable in a digital environment, a Zero-Order Hold (ZOH) discretization is performed:

\begin{equation} \tilde{A} = \exp(A \Delta t), \quad \tilde{B} = (\Delta t A)^{-1} (\exp(A \Delta t) - I) B \end{equation}

The state update equation at each time step is:

\begin{equation} h_t = \tilde{A} h_{t-1} + \tilde{B} x_t \end{equation}

This preserves the temporal dynamics of the system in discrete form. Subsequently, a structured convolutional kernel $\tilde{K}$ is used to generate an output sequence:

\begin{equation} \tilde{K}_d = (C\tilde{B}, C\tilde{A}\tilde{B}, \dots, C\tilde{A}^{M-1}\tilde{B}), y_d = x \ast \tilde{K}_d\ \end{equation}

where $d \in \{\text{fwd}, \text{bwd}\}$. This convolutional kernel $\tilde{K}_d$
encapsulates temporal dependencies across multiple time steps and generates the sequence output 
$\tilde{y}_d$ by convolving the input $\tilde{x}$.

The bidirectional output is then calculated by combining forward and backward sequences as:

\begin{equation}y_{\text{bi}} = \frac{1}{2} (y_{\text{fwd}} + y_{\text{bwd}})\end{equation}

This ensures comprehensive temporal context from both directions.
\begin{algorithm}[H]
    \caption{Bidirectional Mamba Block Process}\label{alg:alg1}
    \begin{algorithmic}[1]
        \Require $x_t$: Input sequence of shape $(B, L, D)$
        \Ensure $y_{\text{bi}}$: Combined output sequence of shape $(B, L, D)$
        \State /* Apply Layer Normalization (LN) */ %
        \State $x' \gets \text{LN}(x_t)$
        \State /* Apply Linear Projection (LP) */ %
        \State $x \gets \text{Linear}(x_t)$ 
        \State $z \gets \text{Linear}(x_t)$ 
        \State /* Apply Mamba Module */ %
        \For{each direction $d \in \{\text{fwd}, \text{bwd}\}$}
            \State $x_d \gets \text{SiLU}(\text{Conv1D}_d(x'_{\text{proj}}))$
            \State /* Compute Parameters for SSM */ %
            \State $A_d, B_d, C_d \gets \text{Linear Parameters for SSM}$ 
            \State $\Delta_d \gets \text{Softplus}(\text{Linear}(x_d))$ 
            \State /* Compute Parameters for discretized SSM */
            \State $A'_d \gets \Delta_d \otimes A_d$ 
            \State $B'_d \gets \Delta_d \otimes B_d$ 
            \State /* Apply State Space Model (SSM) */ %
            \State $y_d \gets \text{SSM}(A'_d, B'_d, C_d)(x_d)$
        \EndFor
        
        \State /* Apply Gating to Output */ %
        \State $y_{\text{fwd}} \gets y_{\text{fwd}} \odot \text{SiLU}(z)$
        \State $y_{\text{bwd}} \gets y_{\text{bwd}} \odot \text{SiLU}(z)$ 
        \State /* Residual Connection and Combine Outputs */ %
        \State $y_{\text{bi}} \gets \text{Linear}\left( y_{\text{fwd}} + y_{\text{bwd}} \right) + x_t$ 
        
        \State \Return $y_{\text{bi}}$
    \end{algorithmic}
\end{algorithm}

The Bidirectional Mamba module integrates forward and backward temporal modeling, providing a balanced and robust feature representation that enhances the overall classification accuracy, particularly in the presence of long EEG sequences with complex temporal dynamics.

\subsection{Optimization Techniques and Data Handling}
In the task of sleep stage classification, model performance is often hindered by issues such as data imbalance, optimization instability, and improper learning rate selection. To address these challenges, we propose a series of optimization techniques and data handling methods aimed at enhancing the robustness and generalization of the model. This section introduces three key approaches: class-aware loss function, learning rate adjustment strategies, and imbalanced data handling techniques.
\subsubsection{Balanced Loss and Data Resampling}
In sleep stage classification, the significant class imbalance often skews model performance toward majority classes, making it essential to implement strategies that directly address this issue. To mitigate this imbalance and improve model generalization, we adopted a combined approach using a class-aware loss function alongside advanced data resampling techniques.

The Focal Loss proposed by Tsung-Yi Lin et al. in their seminal work, Focal Loss for Dense Object Detection \cite{focalloss}, was initially designed for binary classification tasks. It dynamically adjusts the contribution of each sample to the overall loss based on the prediction confidence, effectively reducing the impact of well-classified instances and forcing the model to focus on harder-to-classify samples. The original Focal Loss formula for binary classification is defined as:
\begin{equation}
\text{Loss}(x) = -\alpha (1 - p_t)^{\gamma} \log(p_t)
\end{equation}
where \( p_t = \text{sigmoid}(x) \) represents the probability of the correct class, \(\alpha\)
is a weighting factor that balances the importance of different classes, and \(\gamma\) is a focusing parameter that modulates the loss contribution of well-classified samples.

For our task in multi-class sleep stage classification, we extended this binary Focal Loss to a multi-class formulation. This extension involves applying the Focal Loss across multiple classes by summing over each class-specific loss component. The multi-class Focal Loss formula we employed is given by:
\begin{equation}
\text{Loss}(y, \hat{p}) = -\frac{1}{N} \sum_{i=1}^N \sum_{c=1}^C \alpha_c (1 - \hat{p}_{i,c})^{\gamma} \log(\hat{p}_{i,c})
\end{equation}
where \( N \) is the total number of samples, \( C \) denotes the number of classes, and \( \hat{p}_{i,c} \) is the predicted probability for sample \( i \) belonging to class \( c \), computed as:
\begin{equation}
\hat{p}_{i,c} = \frac{e^{x_{i,c}}}{\sum_{j=1}^{C} e^{x_{i,j}}}
\end{equation}

Here, \( \alpha_c \) represents a class-specific weight that adjusts each class’s contribution to the overall loss, allowing for an increased focus on minority classes in imbalanced datasets. The parameter \( \gamma \) is a focusing parameter that reduces the loss contribution from well-classified samples, directing the model’s attention toward harder-to-classify instances. This softmax-based multi-class Focal Loss effectively normalizes across the probability distribution of all classes, making it well-suited for multi-class tasks and handling mutual exclusivity among classes.

To further refine data distribution, we incorporated Borderline SMOTE, a technique that generates synthetic samples specifically within the boundary regions of minority classes, where misclassification is most likely. By amplifying minority class instances near decision boundaries, Borderline SMOTE enhances the model's sensitivity to difficult-to-classify samples and strengthens overall classification robustness. This approach provides greater representation for minority classes, effectively addressing class imbalance and boosting model performance.

After addressing class imbalance with a balanced loss function and data resampling techniques, it is equally important to ensure stable model training through appropriate learning rate adjustments.

\subsubsection{Learning Rate Adjustment} 
To ensure stable optimization and prevent the model from getting stuck in local minima, we designed an adaptive learning rate scheduling strategy that combines the benefits of ReduceLROnPlateau (RRP) and Cyclic Learning Rate (CLR) techniques, with an additional safeguard of early stopping.

To avoid overfitting or slow convergence when performance plateaus, we employed RRP to dynamically reduce the learning rate based on validation performance. This strategy ensures that the model continues to fine-tune parameters with a lower learning rate after initial rapid progress, helping to stabilize the optimization process.

In summary, the integration of the Temporal Enhancement Block (TEB) and Bidirectional Mamba provides a powerful framework for capturing detailed temporal dependencies in EEG signals. To further optimize the Mamba Mechanism, especially for sleep stage classification, addressing imbalanced data distribution is crucial. By incorporating Balanced Loss and Data Resampling techniques, we ensure the model remains sensitive to underrepresented sleep stages, resulting in more accurate and balanced classification across all stages.

\section{Experiments}
In this section, we conduct comprehensive experiments to evaluate the performance of the proposed BiT-MamSleep architecture.We present an analysis based on multiple publicly available datasets and compare BiT-MamSleep to state-of-the-art approaches. In addition, we conduct an ablation study to assess the contributions of individual modules and optimization strategies within BiT-MamSleep.
\subsection{Datasets and Preprocessing}
To rigorously evaluate the performance of BiT-MamSleep, we employed four widely-used public sleep stage classification datasets: Sleep-EDF-20, Sleep-EDF-78, the Sleep Heart Health Study (SHHS), and ISRUC-S3. To ensure experimental consistency across different datasets, we standardized our input to a single EEG channel for each dataset.

\subsubsection{Sleep-EDF-20 and Sleep-EDF-78} The Sleep-EDF-20 and Sleep-EDF-78 datasets were obtained from PhysioBank \cite{EDF}. Sleep-EDF-20 contains data from 20 healthy subjects aged 25 to 34, recorded during two consecutive nights as part of the Sleep Cassette (SC) study, which focused on the effects of aging on sleep. Sleep-EDF-78, an expanded version, includes 78 recordings from both the SC and Sleep Telemetry (ST) studies. While SC investigates aging effects, the ST study examines the impact of temazepam on sleep in 22 Caucasian participants. Each polysomnographic (PSG) file contains two EEG channels (Fpz-Cz, Pz-Oz), one EOG channel, and one chin EMG channel, all sampled at 100 Hz. For consistency with previous studies \cite{EDF_pre1}, \cite{EDF_pre2}, \cite{EDF_pre3}, we used only the Fpz-Cz channel as input in our experiments.

\subsubsection{Sleep Heart Health Study (SHHS)} The SHHS dataset, a multi-center cohort study, focuses on the cardiovascular and other health consequences of sleep-disordered breathing \cite{SHHS}. From the original 6,441 participants, we selected 329 subjects with regular sleep patterns to minimize the influence of comorbidities such as lung and cardiovascular diseases \cite{SHHS_pre1}. For our experiments, we used the C4-A1 EEG channel, originally sampled at 125 Hz and downsampled to 100 Hz for consistency with other datasets. The recordings were segmented into 30-second epochs in accordance with American Academy of Sleep Medicine (AASM) guidelines.

\subsubsection{ISRUC-S3 Dataset}
The ISRUC-S3 dataset represents a crucial advancement in the field of sleep stage classification, offering an extensive collection of PSG recordings from individuals with potential sleep disorders. Unlike traditional datasets, ISRUC-S3 provides high-resolution, clinically annotated data across multiple channels, including EEG, EOG, and EMG, recorded overnight and segmented into the standard 30-second epochs. The annotations, meticulously labeled by certified sleep technicians following the AASM guidelines, encompass the five primary sleep stages (W, N1, N2, N3, and REM), ensuring robust support for multi-stage classification tasks \cite{ISRUC_Sleep}.This dataset was sampled at 200 Hz, and we selected the C3-A2 channel as input for the model. Although this differs from the channels used in the other datasets, the choice of C3-A2 for ISRUC-S3 aligns with previous studies and maximizes the data quality available in this dataset.

\begin{table}[ht]
\centering
\caption{Dataset Statistics for Sleep Staging}
\resizebox{\columnwidth}{!}{%
\begin{tabular}{lcccccccc}
\toprule
\textbf{Datasets} & \textbf{Channel} & \textbf{W} & \textbf{N1} & \textbf{N2} & \textbf{N3} & \textbf{REM} & \textbf{\#Total} \\
\midrule
Sleep-EDF-20 & Fpz-Cz & 8285 & 2804 & 17799 & 5703 & 7717 & 42308 \\
 & & & \small{19.6\%} & \small{6.6\%} & \small{42.1\%} & \small{13.5\%} & \small{18.2\%} & \\
\midrule
Sleep-EDF-78 & Fpz-Cz & 65951 & 21522 & 69132 & 12029 & 25835 & 195479 \\
 & & & \small{33.7\%} & \small{11.0\%} & \small{35.4\%} & \small{6.7\%} & \small{13.2\%} & \\
\midrule
SHHS & C4-A1 & 4741 & 783 & 8204 & 2747 & 3546 & 20021 \\
 & & & \small{23.7\%} & \small{3.9\%} & \small{40.9\%} & \small{13.7\%} & \small{17.8\%} & \\
\midrule
ISRUC & C3-A2 & 1674 & 1217 & 2616 & 2016 & 1066 & 8589 \\
 & & & \small{19.5\%} & \small{14.2\%} & \small{30.5\%} & \small{23.5\%} & \small{12.4\%} & \\
\bottomrule
\end{tabular}%
}
\end{table}

For all the datasets, we applied a consistent preprocessing pipeline. First, any UNKNOWN stages, which did not align with valid sleep stages, were removed. Second, following AASM guidelines, we combined the N3 and N4 stages into a single N3 stage. Lastly, to emphasize the transitions between sleep stages, we included 30 minutes of wakefulness before and after the sleep periods.

\begin{table*}[ht]
\centering
\renewcommand{\arraystretch}{1.15}
\caption{Comparison Among AttnSleep and State-of-the-Art Models. The Best Values on Each Dataset Are Highlighted in Bold.}
\begin{tabularx}{7in}{c||c|*{5}{X}|*{4}{X}}
\toprule
\multirow{2}{*}{Dataset} & \multirow{2}{*}{Method} & \multicolumn{5}{c|}{Per-Class F1-score} & \multicolumn{4}{c}{Overall Metrics} \\
 \cline{4-6}
 \cline{9-10}
 & & W & N1 & N2 & N3 & REM & Accuracy & MF1 & $\kappa$ & MGm \\
\midrule
\multirow{5}{*}{Sleep-EDF-20} 
  & RandomForest \cite{rf} & 32.13 & 0.00 & 66.96 & 59.81 & 23.20 & 53.49 & 36.42 & 28.19 & 56.62 \\
  & DeepSleepNet \cite{ref13} & 90.48 & 37.5 & 88.38 & 90.47 & 78.24 & 84.09 & 77.02 & 78.06 & 85.71 \\
  & ResnetLSTM \cite{res} & 90.33 & 26.31 & 88.50 & 86.74 & 76.97 & 82.88 & 73.77 & 77.11 & 83.65 \\
  & AttnSleep \cite{ref16} & 88.38 & 40.24 & 88.14 & \textbf{90.33} & 78.49 & 83.66 & 77.1 & 77.54 & 85.57 \\
  & BiT-MamSleep (ours) & \textbf{91.00} & \textbf{41.47} & \textbf{89.84} & 90.08 & \textbf{80.15} & \textbf{85.18} & \textbf{78.51} & \textbf{79.59} & \textbf{86.74} \\
\midrule
\multirow{5}{*}{Sleep-EDF-78} 
  & RandomForest \cite{rf} & 63.61 & 4.37 & 61.04 & 2.12 & 7.89 & 52.43 & 27.81 & 28.07 & 51.91 \\
  & DeepSleepNet \cite{ref13} & 91.35 & 29.33 & 83.38 & 80.29 & 68.52 & 79.41 & 70.57 & 71.26 & 81.55 \\
  & ResnetLSTM \cite{res} & 91.38 & 32.19 & 82.66 & 79.51 & 67.01 & 78.53 & 70.55 & 70.23 & 81.53 \\
  & AttnSleep \cite{ref16} & 91.60 & \textbf{43.69} & 83.45 & \textbf{80.96} & 71.98 & 79.88 & 74.37 & 72.44 & \textbf{83.82} \\
  & BiT-MamSleep (ours) & \textbf{91.70} & 37.71 & \textbf{83.83} & 80.94 & \textbf{72.18} & \textbf{80.21} & \textbf{74.37} & \textbf{72.68} & 83.52 \\
\midrule
\multirow{5}{*}{SHHS} 
  & RandomForest \cite{rf} & 29.40 & 0 & 70.66 & 69.51 & 46.03 & 61.55 & 43.12 & 40.1 & 60.85 \\  
  & DeepSleepNet \cite{ref13} & 85.4 & 40.5 & 77.9 & 81.0 & 73.9 & 81.0 & 73.9 & 0.73 & 81.0 \\
  & ResnetLSTM \cite{res} & 85.1 & 29.4 & 78.9 & 82.6 & 74.6 & 84.2 & 74.6 & 0.76 & 84.2 \\
  & AttnSleep \cite{ref16}  & 85.82 & \textbf{31.91} & 86.32 & \textbf{86.70} & \textbf{80.84} & 83.07 & 74.32 & 76.35 & \textbf{84.69} \\ 
  & BiT-MamSleep (ours) & \textbf{88.56} & 31.83 & \textbf{86.33} & 86.36 & 80.48 & \textbf{84.38} & \textbf{74.71} & \textbf{77.15} & 84.25 \\
\midrule  
\multirow{5}{*}{ISRUC-S3} 
  & RandomForest \cite{rf} & 11.30 & 26.20 & 52.96 & 70.56 & 5.20 & 46.00 & 46.00 & 26.75 & 56.07 \\
  & DeepSleepNet \cite{ref13} & 84.45 & 51.54 & 75.53 & 84.35 & 64.91 & 73.73 & 72.16 & 64.65 & 81.38 \\
  & ResnetLSTM \cite{res} & 85.58 & 46.63 & 74.45 & 83.69 & 63.66 & 73.27 & 70.79 & 63.76 & 82.00 \\
  & AttnSleep \cite{ref16} & 82.75 & \textbf{52.49} & \textbf{78.06} & 84.00 & \textbf{67.82} & 74.22 & \textbf{73.02} & \textbf{66.76} & 82.25 \\ 
  & BiT-MamSleep (ours) & \textbf{87.43} & 50.17 & 76.17 & \textbf{85.18} & 64.52 & \textbf{75.22} & 72.66 & 68.14 & \textbf{82.36 }\\ 
\bottomrule
\end{tabularx}
\end{table*}

\subsection{Baselines and Experimental Setup} \label{sec:experimental_setup}
We compared our against several widely-used traditional machine learning and deep learning models for sleep stage classification. Below, we provide an overview of each baseline method, including its core mechanism and application to sleep staging.
\begin{itemize}

\item{Random Forest \cite{rf}: Random Forests use an ensemble of decision trees to classify EEG data, offering robustness against overfitting and noisy features. However, this method struggles with imbalanced distributions and high-dimensional EEG datasets, making it a limited baseline when compared with deep learning approaches that capture complex patterns.}

\item{DeepSleepNet \cite{ref13}: Combines CNNs for feature extraction and BiLSTM for temporal modeling, effective for sleep stage classification but limited in handling long-range dependencies compared to attention-based models.}

\item{ResnetLSTM \cite{res}: Integrates ResNet for robust feature extraction with LSTM for temporal transitions, providing strong spatial-temporal modeling but struggles with very long-range dependencies.}

\item{AttnSleep \cite{ref16}: Introduces multi-head self-attention for capturing long-range dependencies in EEG sequences, outperforming RNN-based models in processing large-scale and complex temporal relationships.}

\end{itemize}

We implemented our BiT-MamSleep model using the PyTorch framework (v2.1.2) on an Ubuntu 22.04 system with CUDA 11.8. The model was trained using the Adam optimizer with an initial learning rate of 
\(\eta = 10^{-3}\). Training continued for up to 100 epochs with a batch size of 128. A subject-wise 5-fold cross-validation was applied across all the datasets we introduced before to ensure robust evaluation. 

To address class imbalance, we employed Focal Loss, particularly focusing on the underrepresented sleep stages (e.g., N1 and REM). We incorporated an early stopping strategy that halted training if the validation accuracy did not improve after 30 epochs. If there was no improvement for 15 consecutive epochs, we switched from the RRP scheduler to a CyclicLR scheduler. This induced periodic learning rate oscillations to help the model escape local minima.


\subsection{Experiment Results}
In this section, we evaluate the performance of the proposed BiT-MamSleep model across various datasets, including EDF-20, EDF-78, and SHHS. Sleep stage classification is inherently challenging due to the imbalanced nature of the datasets, where certain stages (e.g., N1 and REM) are underrepresented. To ensure robust and fair evaluation, we adopted four well-established metrics: accuracy (ACC), macro-averaged F1-score (MF1), Cohen's Kappa ($\kappa$) \cite{cohen1960}, and the macro-averaged G-mean (MGm). Each of these metrics provides unique insights into the performance of the model across all sleep stages, particularly focusing on both the dominant and underrepresented classes.

\begin{itemize}
\item \textbf{Accuracy (ACC)}: This metric measures the overall proportion of correctly classified sleep stages in the dataset. It is defined as:
\begin{equation}
    \text{ACC} = \frac{TP + TN}{TP + TN + FP + FN}
\end{equation}
where $TP$, $TN$, $FP$, and $FN$ represent the true positives, true negatives, false positives, and false negatives, respectively.

\item \textbf{Macro-averaged F1-score (MF1)}: This evaluates the harmonic mean of precision and recall across all sleep stages, giving equal weight to each class, which is crucial for addressing class imbalance. It is calculated as:
\begin{equation}
    \text{MF1} = \frac{1}{C} \sum_{i=1}^{C} \text{F1}_i
\end{equation}
where $C$ is the total number of classes, and $\text{F1}_i$ is the F1-score for class $i$.

\item \textbf{Cohen's Kappa ($\kappa$)}: This metric adjusts for chance agreement between predicted and actual classes, providing a more realistic evaluation of the model's classification performance. It is defined as:
\begin{equation}
    \kappa = \frac{p_o - p_e}{1 - p_e}
\end{equation}
where $p_o$ is the observed agreement and $p_e$ is the expected agreement by chance.

\item \textbf{Macro-averaged G-mean (MGm)}: This measures the geometric mean of sensitivity across all classes, further highlighting the model's ability to maintain balanced performance across underrepresented sleep stages. It is given by:
\begin{equation}
    \text{MGm} = \left( \prod_{i=1}^{C} \text{Sensitivity}_i \right)^{\frac{1}{C}}
\end{equation}
where $\text{Sensitivity}_i$ represents the sensitivity (recall) for class $i$, and $C$ is the number of classes.
\end{itemize}

The BiT-MamSleep was extensively evaluated across four well-established sleep datasets, Sleep-EDF-20, Sleep-EDF-78, SHHS, and ISRUC-S3, to rigorously assess its capacity for accurate, robust, and balanced sleep stage classification. Each dataset presents unique challenges, from class imbalance to diverse EEG patterns, enabling a comprehensive comparison with aforementioned baseline models: Random Forest, DeepSleepNet, ResNetLSTM, and AttnSleep. The results reveal that BiT-MamSleep significantly outperforms these models in both overall accuracy and in handling critical classification challenges, including long-range temporal dependencies, multi-scale feature extraction, and imbalanced sleep stage data.

On the Sleep-EDF-20 dataset, BiT-MamSleep achieved an accuracy of 85.18\%, outperforming established models by at least 2\% and showcasing superior robustness with a macro-averaged F1-score of 78.51\% and a Cohen’s Kappa of 79.59\%. This dataset’s imbalanced distribution, where common stages like N2 dominate over less frequent ones like N1 and REM, which pose substantial challenges to baseline models. For instance, DeepSleepNet, which combines CNN and LSTM layers, exhibited limitations in capturing the temporal transitions necessary for accurately detecting minority stages. Similarly, ResNetLSTM, though designed to capture temporal patterns, struggled with the dataset's long EEG sequences due to gradient-related issues inherent in recurrent architectures, often resulting in reduced accuracy on rare stages. AttnSleep, which employs attention to capture temporal dependencies, performed better overall but is constrained by its unidirectional processing, limiting its ability to fully capture bidirectional context in stage transitions.

In contrast, BiT-MamSleep addresses these limitations with its BiMamba mechanism, which enables the model to learn both forward and backward dependencies, crucial for accurately modeling sleep stage transitions over extended EEG sequences. This dual-directional processing proved particularly advantageous in identifying stages like N1 and REM, where subtle temporal dependencies are essential. Furthermore, BiT-MamSleep’s adaptive feature recalibration dynamically adjusts the relevance of specific features across stages, allowing it to effectively classify even the minority stages, thereby enhancing the model’s balanced performance across all classes.

On the Sleep-EDF-78 dataset, BiT-MamSleep maintained its edge, achieving an accuracy of 80.21\%. This dataset, characterized by a broader subject base and greater EEG variability, challenged baseline models to generalize effectively. While AttnSleep’s self-attention mechanism improved its performance over recurrent models by capturing longer dependencies, its single-pass processing lacked the dual temporal context provided by BiT-MamSleep’s bidirectional mechanism. BiT-MamSleep’s TRCNN played a crucial role here, enabling multi-resolution feature extraction that effectively captures both high-frequency details and long-range dependencies in EEG signals. This multi-resolution approach allowed BiT-MamSleep to have good performance in the classification of minority stages, such as N1, where previous models showed weaker performance. The adaptability of BiT-MamSleep’s architecture to capture both spatial and temporal characteristics ensured consistent gains over prior models.

BiT-MamSleep further demonstrated its adaptability on the SHHS dataset, achieving an accuracy of 84.38\%. This dataset, which includes data from subjects with sleep-disordered breathing, introduces unique EEG patterns that increase classification complexity. Random Forest, as a non-deep-learning baseline, struggled with these challenges due to its reliance on hand-crafted features, which do not capture the intricate temporal patterns inherent in EEG. Both DeepSleepNet and ResNetLSTM encountered limitations in handling the distinct temporal dependencies introduced by disordered breathing events, as their sequential modeling approaches lack the bidirectional perspective essential for comprehensive temporal coverage. BiT-MamSleep’s bidirectional mechanism again proved instrumental, allowing the model to process both historical and future context in parallel, crucial for capturing the physiological nuances that distinguish each sleep stage. Additionally, the model’s balanced performance, reflected in a high macro-averaged G-mean, confirms its ability to classify each stage accurately, avoiding the skew often seen in models not equipped to handle imbalanced classes.

On the ISRUC-S3 dataset, which provides high-resolution polysomnographic data from patients with potential sleep disorders, BiT-MamSleep achieved 75.22\% accuracy, outperforming all baselines. This dataset’s complex nature, with fine-grained annotations across EEG, EOG, and EMG channels, necessitates multi-resolution processing to identify sleep stages accurately. AttnSleep’s single-resolution attention model lacked the capacity to capture the diverse spectral and temporal features, particularly in stages like REM that require both long-range dependencies and high-frequency detail. BiT-MamSleep’s TRCNN architecture, by employing small, medium, and large convolutional kernels, effectively captured a wide spectrum of EEG frequencies, ensuring reliable classification even in high-resolution and diverse data. The adaptive feature recalibration module further strengthened BiT-MamSleep’s capacity to adapt to the variability of disordered sleep data, dynamically prioritizing stage-specific features, thus ensuring a balanced performance across the five main stages.

Across all datasets, BiT-MamSleep consistently demonstrated its superiority in addressing the limitations of traditional and recent deep learning models through the following advancements:

\begin{itemize}
    \item \textbf{Bidirectional Temporal Modeling:} The BiMamba mechanism’s capacity to capture both forward and backward dependencies effectively overcomes the unidirectional limitations of models like AttnSleep and DeepSleepNet, which struggle with stage transitions and lack temporal robustness.
    \item \textbf{Multi-Resolution Feature Extraction:} The TRCNN architecture’s ability to capture frequency-specific features at various temporal scales enables BiT-MamSleep to handle the non-stationary nature of EEG data, providing it an advantage over both Random Forest and single-resolution deep learning models.
    \item \textbf{Dynamic Feature Recalibration:} By adapting feature importance dynamically, BiT-MamSleep enhances its performance on minority classes, addressing class imbalance challenges more effectively than models like ResNetLSTM, which lack this flexibility.
    \item \textbf{Effective Class Imbalance Handling:} With the use of focal loss and SMOTE-based resampling, BiT-MamSleep achieves balanced sensitivity across all stages, significantly improving accuracy for minority classes like N1 and REM, where traditional models often fall short.
\end{itemize}

\subsection{Ablation Study} 

\begin{figure}[!t]
\centering
\includegraphics[width=3.5in]{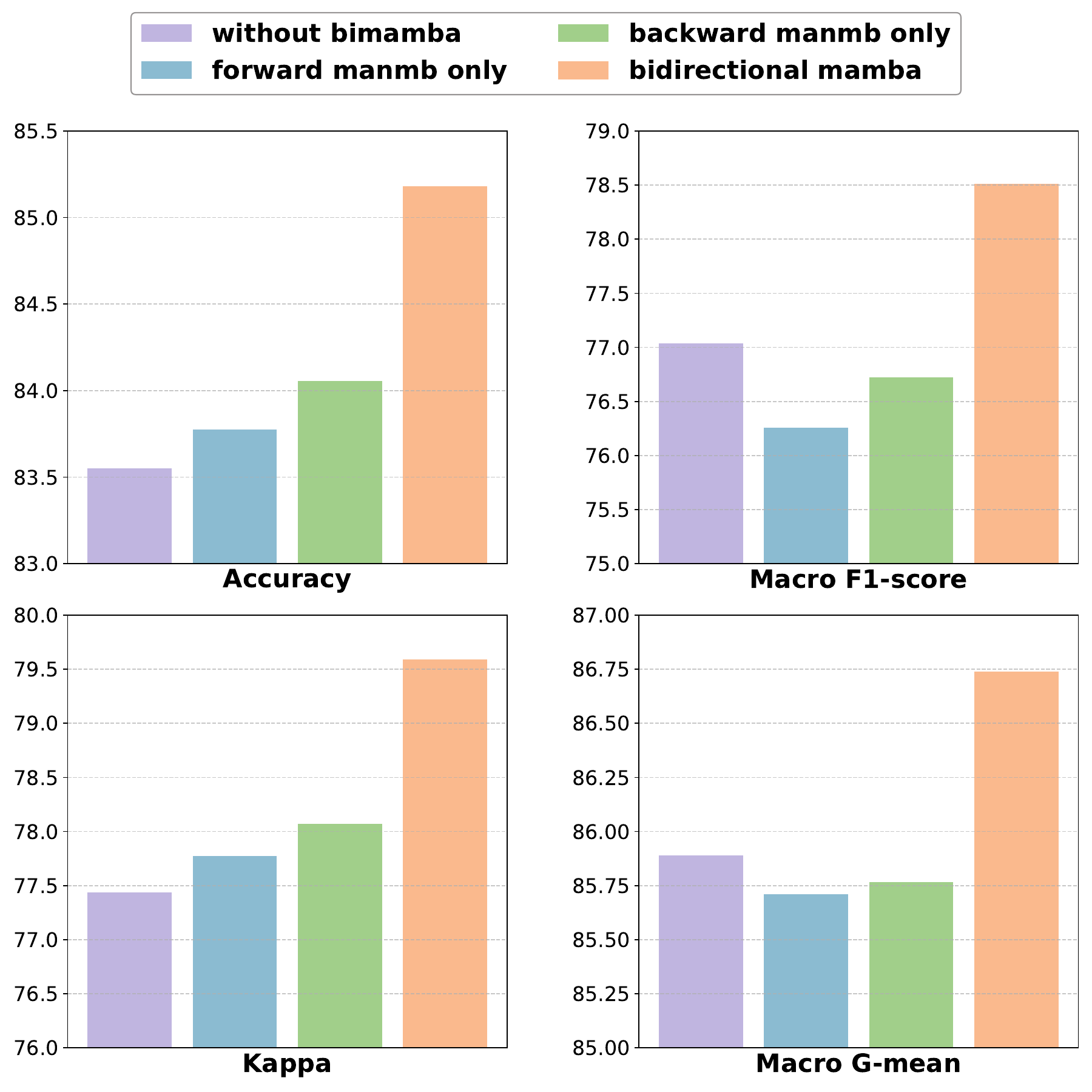}
\caption{Ablation study conducted on Sleep-EDF-20 dataset.}
\label{fig_3}
\end{figure}

The ablation study focuses on evaluating the effectiveness of the Mamba module, the core innovation in the proposed MamSleep architecture. We specifically examine how different configurations of the Mamba module impact the model's performance on sleep stage classification tasks. This study aims to highlight the significance of bidirectional processing in effectively capturing complex temporal dependencies in EEG data.

To evaluate the impact of the Mamba module, we conducted four sets of experiments on the Sleep-EDF-20 dataset using subject-wise 5-fold cross-validation:
\begin{itemize}
\item{Without BiMamba (MRCNN + AFMs): In this setting, the Mamba module was entirely removed, and only the MRCNN along with Adaptive Feature Modules (AFMs) were used for feature extraction and recalibration. This setup serves as a baseline to understand the contribution of the Mamba module.}
\item{Forward Mamba Only: In this configuration, only the forward Mamba was used, capturing temporal dependencies in the forward direction. This helps in assessing the effectiveness of modeling only the past information.}
\item{Backward Mamba Only: Here, only the backward Mamba was employed, which models temporal dependencies in the reverse direction, focusing on future information.}
\item{BiMamba: This is the full MamSleep model, where both forward and backward temporal dependencies are modeled simultaneously, providing a comprehensive representation of EEG signals.}
\end{itemize}

Each of these experiments was evaluated using common metrics including ACC, MF1, and $\kappa$. The results of the ablation experiments are summarized in Table 4.

The ablation study results demonstrate that the BiMamba mechanism significantly enhances the BiT-MamSleep's performance, outperforming other configurations. Without BiMamba, the model's accuracy, MF1, and Cohen's Kappa values decreased, highlighting the limitations of capturing only spatial features. Using only forward Mamba showed some improvement over the baseline (TRCNN + AFMs), indicating the value of capturing past dependencies. However, it still lagged behind BiMamba, as forward modeling alone is insufficient for capturing complete temporal context needed for accurate sleep stage transitions. Similarly, using only backward Mamba improved performance compared to the baseline but was still inferior to the full BiMamba model, due to the lack of past context. The BiMamba's bidirectional approach effectively captures both past and future dependencies, leading to substantial improvements, particularly in handling imbalanced classes and subtle sleep stage transitions, making it the most effective approach for accurate sleep stage classification.

\section{Conclusion}
In this paper, we presented MamSleep, a novel architecture designed for automatic sleep stage classification using EEG data. Our model integrates the TRCNN for efficient multi-scale feature extraction and the BiMamba mechanism, which effectively captures both short- and long-term temporal dependencies through bidirectional processing. The experimental results on multiple public datasets demonstrate that MamSleep significantly outperforms state-of-the-art methods in terms of accuracy, F1-score, and Kappa, particularly excelling in handling long EEG sequence.

The core innovation, the BiMamba mechanism, proved to be crucial in modeling complex temporal dependencies inherent in sleep EEG data, allowing for improved recognition of sleep stage transitions. The integration of adaptive feature recalibration and temporal enhancement blocks further refined feature representation, contributing to overall classification robustness and efficiency.

While BiT-MamSleep demonstrates significant improvements, future work could explore optimizing the computational complexity of the model, making it more suitable for real-time applications such as wearable devices for sleep monitoring. Additionally, we plan to evaluate the generalizability of BiT-MamSleep on other physiological signal datasets and investigate its potential for broader health monitoring applications. By advancing EEG-based sleep classification, BiT-MamSleep aims to contribute to more accessible and accurate sleep health assessments in clinical and consumer settings.


\end{document}